\documentclass[a4paper,11pt]{article}
\UseRawInputEncoding
\pdfoutput=1 

\usepackage [latin1]{inputenc}
\usepackage{jheppub} 

\usepackage{amsmath,amssymb,graphics,epsfig,subfigure}

\title{\boldmath Dynamic Property of Phase Transition for Non-Linear Charged Anti-de Sitter black holes}

\author[]{Yun-Zhi Du,}
\author[]{Huai-Fan Li \footnote{Corresponding author}}
\author[]{Fang Liu,}
\author[]{Li-Chun Zhang}

\affiliation[]{Department of Physics, Shan xi Da tong University, Da tong 037009, China\\
Institute of Theoretical Physics, Shan xi Da tong University, Da tong 037009, China}
\emailAdd{duyzh13@lzu.edu.cn}
\emailAdd{huaifan999@163.com}
\emailAdd{sapphire513@126.com}
\emailAdd{zhlc2969@163.cn}

\abstract{Understanding the thermodynamic phase transition of black holes can provide the deep insight into the fundamental properties of black hole gravity to establish the quantum gravity. In this work, we investigate the phase transition and its dynamics for the charged EPYM AdS black hole. Through reconstructing the Maxwell's equal-area law, we find there exists the high-/low-potential black hole (HPBH/LPBL) phase transition, not only the pure large/small one. The Gibbs free energy landscape ($G_L$) is treated as a function of the black hole horizon, which is the order parameter of the phase transition due to the thermal fluctuation. From the view point of $G_L$, the stable HPBH/LPBL states are corresponding to two wells of $G_L$, which have the same depth. The unstable intermediate-potential black hole state corresponds to the local maximum of $G_L$. Then we focus on the probability evolution governed by the Fokker-Planck equation. Through solving the Fokker-Planck equating with different reflection/aborption boundary conditions and initial conditions, the dynamics of switching between the coexistent HPBH and LPBL phases is probed within the first passage time. Furthermore, the effect of temperature on the dynamic property of phase transition is also investigated.}


\begin{document}
\maketitle
\flushbottom

\section{Introduction}

In $1983$, the Hawking-Page (HP) phase transition of an AdS spacetime was proposed by Hawking and Page, which could give the evolution of spacetime with different phases \cite{Hawking1983}. Namely, with increasing temperature the dominant configuration is from the pure thermal radiation phase, then to the coexistent phase with an AdS black hole and thermal radiation, and finally to a stable black hole. Witten had explained it as a confinement/deconfinement phase transition in gauge theory in Ref. \cite{Witten1998}. And it could also be understood as a solid/liquid phase transition \cite{Altamirano2013} by regarding the cosmological constant as pressure $P=-\frac{\Lambda}{8\pi}=\frac{(n-1)(n-2)}{16\pi l^2}$, whose conjugate variable is the thermodynamic volume. Subsequently the phase transition in the extended AdS/dS phase space has been widely considered \cite{Hendi2017a,Hennigar2017a,Frassin,Kubiznak2012,Cai2013,Ma2017,Ma2017a,Mir2017b,Banerjee2017,Banerjee2011,Hendi2019,Simovic2019,Hennigar2019,Mbarek2019,
Kubiznak2016,Li2017a,Li2017,zhao2015,Ma2018,Ma2016,Zhang2016,Zhou2019,Dinsmore2020,Cai1306,Wei1209,Caldarelli2000,Mann1207,Wei2015,Banerjee1109,Hendi1702,
Bhattacharya2017,Zeng2017,Hendi1803,Zhang1502,Cheng1603,Zou1702,Dolan2014,Altamirano2014,Du2021,Du2021a,Zhang2020}.

As we know, the phase transition of an ordinary thermodynamic system is a result of the competition between the micro-components. Since the black holes have the similar thermal behavior as an ordinary thermodynamic system, the microstructure of black hole becomes a hit issue. The authors in Ref. \cite{Wei2015} investigated the AdS black hole microstructure by the Ruppeiner scalar curvature following the Ruppeiner geometry \cite{Rupperiner1996}. The number density of the speculative black hole molecules was introduced to examine its phase transition and microstructure. The way of indicating different kinds of interaction between black hole molecules by the different values of the Ruppeiner scalar curvature was quickly generalized to other black holes \cite{Ruppeiner2014,Ruppeiner2018,Ruppeiner2008,Ruppeiner2012,Miao2019,Miao2017}.

With the great development of phase transition and microstructure for various AdS black holes, people attempt to probe the dynamic process of black hole phase transition. Recently from the view point of the Gibbs free energy landscape ($G_L$), the authors in Ref. \cite{Li2020} probed the dynamics of switching between the coexistence black hole phases by solving the Fokker-Planck equating with different reflection/aborption boundary conditions and initial conditions, and calculating the mean first passage time. In this approach, the phase transition is due to the thermal fluctuation and $G_L$ is regarded an function of black hole horizon which is the order parameter of phase transition. Subsequently, this method was applied to the HP phase transition in Einstein gravity \cite{Li2021} and in massive gravity \cite{Li2020a} and the large/small black hole phase transition in Gauss-Bonnet gravity \cite{Wei2021}, in Einstein gravity \cite{Yang2021} minimally coupled to nonlinear electrodynamics \cite{Kumara2021}, and in dilaton gravity \cite{Mo2021}.

The linear charged black holes in AdS spacetime \cite{Chamblin1999,Chamblin1999a} within a second-order phase transition shows a scaling symmetry: at the critical point the state parameters scale respects to charge q, i.e., $S\sim q^2,~P\sim q^{-2},~T\sim q^{-1}$ \cite{Johnson2018,Johnson2018a}. It is naturally to gauss whether these exists the scaling symmetry in the non-linear charged AdS black holes. As a generalization of the charged AdS Einstein-Maxwell black holes, it is interesting to explore new non-linear charged systems. Due to infinite self-energy of point like charges in Maxwell's theory \cite{Born1934,Kats2007,Anninos2009,Cai2008,Seiberg1999}, Born and Infeld proposed a generalization when the field is strong, bringing in non-linearities \cite{Dirac2013,Birula1970}. An interesting non-linear generalization of charged black holes involves a Yang-Mill field coupled to Einstein gravity, where several features in extended thermodynamics have recently been studied \cite{Du2021,Zhang2015,Moumni2018}.

Inspired by these, we will probe the behaviour of the Gibbs free energy landscape for the charged AdS black holes and explore the dynamic process of the low/high-potential black hole phase transition in the four-dimensional Einstein-power-Yang-Mills (EPYM) gravity. This work is organized as follows. In Sec. \ref{scheme2}, we briefly review the thermodynamic quantities and high-/low-potential black hole (HPBH/LPBL) phase transition of the charged EPYM AdS black hole by the Maxwell's equal-area law. In Sec. \ref{scheme3}, we present the behaviour of the Gibbs free energy landscape at the phase transition point and explore the probability evolution through solving the Fokker-Planck equating with different reflection/aborption boundary conditions and initial conditions. Then the dynamics of switching between the coexistence high-/low-potential black hole states is probed by calculating the mean first passage time. Furthermore, we also investigate the effect of temperature on the dynamic property of phase transition. A brief summary is given in Sec. \ref{scheme4}.

\section{Thermodynamic and Phase Transition for Non-Linear Charged AdS Black Holes}
\label{scheme2}
In this section, we will give a brief review of the thermodynamic and phase transition of the non-linear charged AdS black hole.
\subsection{Thermodynamic quantities}
\label{scheme2.1}
The action for four-dimensional Einstein-power-Yang-Mills (EPYM) gravity with a cosmological constant $\Lambda$ was given by \cite{Zhang2015,Corda2011,Mazharimousavi2009,Lorenci2002}
\begin{eqnarray}
I=\frac{1}{2}\int d^4x\sqrt{g}
\left(R-2\Lambda-[Tr(F^{(a)}_{{\mu\nu}}F^{{(a)\mu\nu}})]^\gamma\right)
\end{eqnarray}
with the Yang-Mills (YM) field
\begin{eqnarray}
F_{\mu \nu}^{(a)}=\partial_{\mu} A_{\nu}^{(a)}-\partial_{\nu} A_{\mu}^{(a)}+\frac{1}{2 \xi} C_{(b)(c)}^{(a)} A_{\mu}^{b} A_{\nu}^{c}
\end{eqnarray}
where $Tr(F^{(a)}_{\mu\nu}F^{(a)\mu\nu})
=\sum^3_{a=1}F^{(a)}_{\mu\nu}F^{(a)\mu\nu}$, $R$ and $\gamma$ are the scalar curvature and a positive real parameter, respectively. And $C_{(b)(c)}^{(a)}$ represents the structure constants of three parameter Lie group G, $\xi$ is the coupling constant, $A_{\mu}^{(a)}$ are the $SO(3)$ gauge group YM potentials.

For this system, the black hole solution with the negative cosmological constant $\Lambda$ of the corresponding field equation is \cite{Yerra2018}:
\begin{eqnarray}
d s^{2}=-f(r) d t^{2}+f^{-1} d r^{2}+r^{2} d \Omega_{2}^{2},\\
f(r)=1-\frac{2 M}{r}-\frac{\Lambda}{3} r^{2}+\frac{\left(2 q^{2}\right)^{\gamma}}{2(4 \gamma-3) r^{4 \gamma-2}},
\end{eqnarray}
where $d\Omega_{2}^{2}$ is the metric on unit $2$-sphere with volume $4\pi$ and $q$ is the YM charge. Note that this solution is valid for the condition of the non-linear YM charge parameter $\gamma\neq0.75$ and the power YM term holds the weak energy condition (WEC) for $\gamma>0$ \cite{Corda2011}. In the extended phase space, $\Lambda$ was interpreted as the thermodynamic pressure $P=-\frac{\Lambda}{8\pi}$. The black hole event horizon locates at $f(r_+)=0$. The parameter $M$ represents the ADM mass of the black hole and it reads
\begin{eqnarray}
H=M(S, q, P)=\frac{1}{6}\left[8 \pi P\left(\frac{S}{\pi}\right)^{3 / 2}+3\left(\frac{S}{\pi}\right)^{\frac{3-4 \gamma}{2}} \frac{\left(2 q^{2}\right)^{\gamma}}{8 \gamma-6}+3 \sqrt{\frac{S}{\pi}}\right].
\end{eqnarray}
And in our set up it is associated with the enthalpy of the system. The black hole temperature, entropy, and volume were given by \cite{Zhang2015}
\begin{eqnarray}
T=\frac{1}{4 \pi r_{+}}\left(1+8 \pi P r_{+}^{2}-\frac{\left(2 q^{2}\right)^{\gamma}}{2 r_{+}^{(4 \gamma-2)}}\right),~~~~~S=\pi r_{+}^{2},~~~~V=\frac{4\pi r_+^3}{3}.    \label{T}
\end{eqnarray}
The YM potential $\Psi$ was given by \cite{Anninos2009}
\begin{eqnarray}
\Psi=\frac{\partial M}{\partial q^{2 \gamma}}=\frac{r_{+}^{3-4 \gamma} 2^{\gamma-2}}{4\gamma-3}.\label{Psi}
\end{eqnarray}
The above thermodynamic quantities satisfies the first law
\begin{eqnarray}
d M=T d S+\Psi d q^{2 \gamma}+V d P.
\end{eqnarray}
While the equation of state $P(V,T)$ for canonical ensemble (fixed YM charge $q$) can be obtained from the expression of temperature as
\begin{eqnarray}
P=\left(\frac{4 \pi}{3 V}\right)^{1 / 3}\left[\frac{T}{2}-\frac{1}{8 \pi}\left(\frac{4 \pi}{3 V}\right)^{1 / 3}+\frac{\left(2 q^{2}\right)^{\gamma}}{16 \pi}\left(\frac{4 \pi}{3 V}\right)^{\frac{1-4\gamma}{3}}\right].\label{PTV}
\end{eqnarray}

\subsection{Equal-Area Law}
\label{scheme2.2}

From Eq. (\ref{PTV}), we know the state equation of the EPYM black hole with fixed YM charge corresponds to the one of an ordinary thermodynamic system and it can be written as $f(T,P,V)=0$. Furthermore the number of particles in the system is unchanged. Through the Maxwell's equal-area law, we can construct the phase transition of the EPYM black hole in $P-V$, $T-S$, and $q^{2\gamma}-\Psi$, respectively. For the same parameters, the phase transition points in $P-V$, $T-S$, and $q^{2\gamma}-\Psi$ are the same as shown in Ref. \cite{Du2021}. In the following we will present the $T-S$ phase diagram and label the phase transition point with $T_0$ and $P_0$.

\begin{figure}[htp]
\centering
\includegraphics[width=0.45\textwidth]{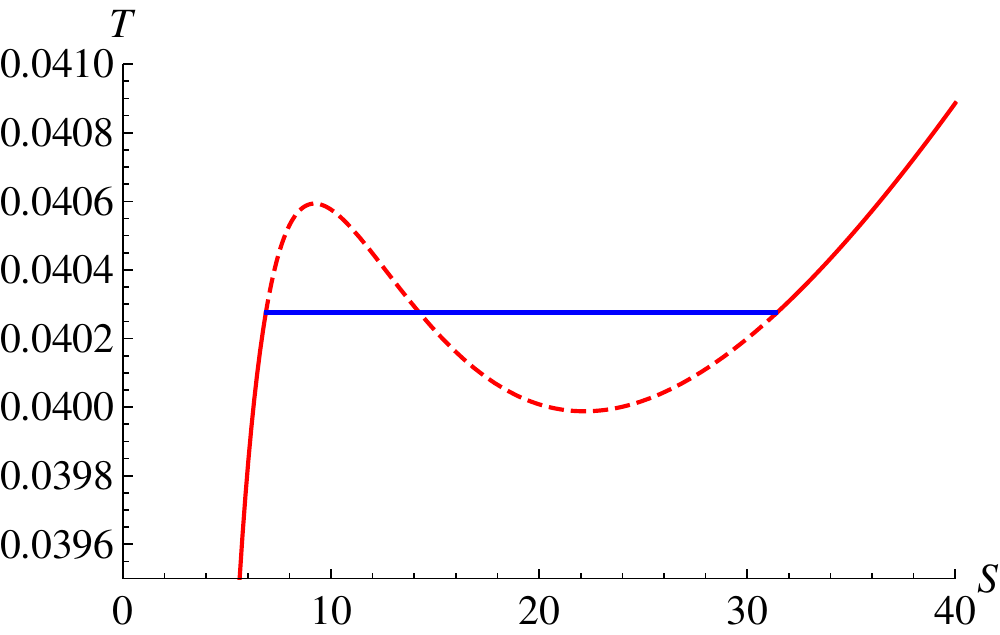}
\caption{The phase transition in $T-S$ diagram with the parameters $q=0.85,~\gamma=0.8$. The corresponding phase transition point is $T_0=0.0402$ and $P_0=0.85955P_c$.}\label{T-S}
\end{figure}

For the EPYM black hole with the given YM charge $q$ and pressure $P_0<P_c$, the entropy at the boundary of the two-phase coexistence area are $S_1$ and $S_2$, respectively. And the corresponding temperature is $T_0$, which is less than the critical temperature $T_c$ and is determined by the horizon radius $r_+$. Therefore, from the Maxwell's equal-area law $T_0(S_2-S_1)=\int^{S_2}_{S_1}TdS$ and Eq. (\ref{T}), we have
\begin{eqnarray}
2 \pi T_{0}=\frac{1}{r_{2}(1+x)}+\frac{8 \pi P_0 r_{2}}{3(1+x)}\left(1+x+x^{2}\right)-\frac{\left(2 q^{2}\right)^{\gamma} r_{2}^{1-4 \gamma}}{2(3-4 \gamma)} \frac{\left(1-x^{3-4 \gamma}\right)}{\left(1-x^{2}\right)}\label{TEAL}
\end{eqnarray}
with $x=\frac{r_1}{r_2}$. In addition, from the state equation we have
\begin{eqnarray}
T_{0}=\frac{1}{4 \pi r_{2}}\left(1+8 \pi P_0 r_{2}^{2}-\frac{\left(2 q^{2}\right)^{\gamma}}{2 r_{2}^{(4 \gamma-2)}}\right),~~~~ \label{T02}\\
T_{0}=\frac{1}{4 \pi r_{1}}\left(1+8 \pi P_0 r_{1}^{2}-\frac{\left(2 q^{2}\right)^{\gamma}}{2 r_{1}^{(4 \gamma-2)}}\right).\label{T01}
\end{eqnarray}
From Eq. (\ref{TEAL}), we have
\begin{eqnarray}
0&=&-\frac{1-x}{r_{2} x}+8 \pi \operatorname{P_0r}_{2}(1-x)+\frac{\left(2 q^{2}\right)^{\gamma}}{2 r_{2}^{4 \gamma-1} x^{4 \gamma-1}}\left(1-x^{4 \gamma-1}\right),\label{T000}~~~~\\
8 \pi T_{0}&=&\frac{1+x}{r_{2} x}+8 \pi \operatorname{P_0r}_{2}(1+x)-\frac{\left(2 q^{2}\right)^{\gamma}}{2 r_{2}^{4 \gamma-1} x^{4 \gamma-1}}\left(1+x^{4 \gamma-1}\right).\label{T001}
\end{eqnarray}
Considering Eqs. (\ref{TEAL}), (\ref{T000}), and (\ref{T001}), we find the larger horizon has the following form
\begin{eqnarray}
r_{2}^{4 \gamma-2}=\frac{\left(2 q^{2}\right)^{\gamma}\left[(3-4 \gamma)(1+x)\left(1-x^{4 \gamma}\right)+8 \gamma x^{2}\left(1-x^{4 \gamma-3}\right)\right]}{2 x^{4 \gamma-2}(3-4 \gamma)(1-x)^{3}}
=\left(2 q^{2}\right)^{\gamma} f(x,\gamma).
\label{r2}
\end{eqnarray}
Since the above state parameters must be positive, the non-linear YM charge parameter satisfies the condition $\frac{1}{2}<\gamma$. For the critical point ($x=1$), the state parameters are
\begin{eqnarray}
r_{c}^{4 \gamma-2}&=&\left(2 q^{2}\right)^{\gamma} f(1, \gamma), \quad f(1, \gamma)=\gamma(4 \gamma-1),\\
T_{c}&=&\frac{1}{\pi\left(2 q^{2}\right)^{\gamma /(4 \gamma-2)} f^{1 /(4 \gamma-2)}(1, \gamma)} \frac{2 \gamma-1}{4 \gamma-1}, \\
P_{c}&=&\frac{2 \gamma-1}{16 \pi \gamma\left(2 q^{2}\right)^{\gamma /(2 \gamma-1)} f^{1 /(2 \gamma-1)}(1, \gamma)}.\label{rcTcPc}
\end{eqnarray}
Substituting Eq. (\ref{r2}) into Eq. (\ref{T000}), we have the following express
\begin{eqnarray}
\frac{1-x}{x}=8\pi P_0 \left(2 q^{2}\right)^{\frac{\gamma}{2 \gamma-1}}f^{\frac{1}{2 \gamma-1}}(x,\gamma)+\frac{1-x^{4 \gamma-1}}{2x^{4 \gamma-1}f(x,\gamma)}f^{\frac{4 \gamma-3}{4 \gamma-2}}(x,\gamma).\label{xchi}
\end{eqnarray}
For the given parameter $\gamma$ and pressure $P_0$, we can obtain the value of $x$ from the above equation. Then from Eq. (\ref{r2}), we know that for the given temperature $P_0$ ($P_0<P_c$), i.e., for the fixed value of $x$, the phase transition condition reads
\begin{eqnarray}
\frac{\left(2 q^{2}\right)^{\gamma}}{r_{2}^{4 \gamma-2}}=\frac{1}{f(x, \gamma)}. \label{PTC}
\end{eqnarray}
Therefore, the phase transition of the EPYM black hole with a given temperature $P_0$ ($P_0<P_c$) is determined by the radio between the YM charge $(2q^2)^\gamma$ and $r_{2}^{4 \gamma-2}$, not only the value of horizon. Note that we call this radio as the YM electric potential with the horizon radius $r_2$. Therefore, the phase transition of this thermodynamic system is the high-/low-potential black hole (HPBH/LPBL) phase transition. The plot of the phase transition in $T-S$ diagram with the fixed pressure $P_0=0.85955P_c$ is shown in Fig. \ref{T-S}. The effects of the non-linear parameter $\gamma$ and YM charge $q$ on phase transition had been exhibited in our last work \cite{Du2021}.

\section{Dynamic Properties of Thermodynamic Phase Transition}
\label{scheme3}
As well as the equal area law, Gibbs free energy is an important thermodynamic quantity to investigate the phase transition and it exhibits a swallow tail behavior at a first-order phase transition point. While it is continuous but not smooth at a second-order phase transition. These two methods are thermodynamically equivalent. Therefore, we do not present the swallow tail behavior in this part.

Recently, authors in Refs. \cite{Li2020a} proposed that Gibbs free energy landscape should also correspond to the thermal dynamic phase transition of a black hole. In this section, we will investigate the thermal dynamic phase transition of the EPYM AdS black hole from the view of Gibbs free energy landscape.
\subsection{Gibbs Free Energy Landscape}
\label{scheme3.1}
For the EPYM charged AdS black hole with non-linear charge, we proposed there exist the HPBH/LPHB phase transitions in Ref. \cite{Du2021}, not only the pure large/small black holes one. In the following we will present the thermal dynamic phase transition at the phase transition point ($T_0=0.0402$ and $P_0=0.85955P_c$) with the parameters $q=0.85,~\gamma=0.8$.

\begin{figure}[htp]
\centering
\includegraphics[width=0.45\textwidth]{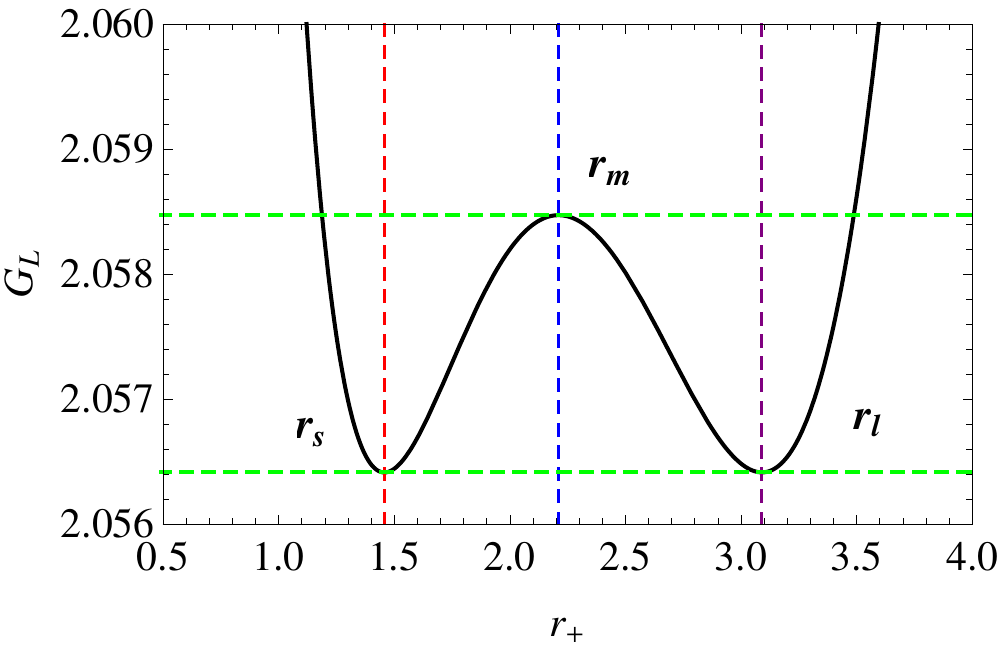}
\caption{The plot of $G_L-r$ at the phase transition point ($T_0=0.0402$ and $P_0=0.85955P_c$) with the parameters $q=0.85,~\gamma=0.8$. }\label{GLr}
\end{figure}

Gibbs free energy landscape of the charge EPYM AdS black hole reads
\begin{eqnarray}
G_L=M-T_{E}S
\end{eqnarray}
where $T_E$ is a temperature parameter and equals $T_0$, which is not the Hawking temperature. The picture of Gibbs free energy landscape at phase transition point of $P_0=0.85955P_c$ and $T_E=T_0=0.0402$ is exhibited in Fig. \ref{GLr}. From this picture, we can see that at the phase transition point the Gibbs free energy landscape displays the double-well behavior. Namely there are two local minimum (located at $r_l=1.4577,~r_s=3.0904$) which are corresponding to the stable HPBH and LPBL states with positive heat capacity. The local maximum located at $r_m=2.2107$ stands for the unstable intermediate-potential black hole state with negative heat capacity and acts as a barrier between the stable HPBH and LPBH states. Furthermore, the depths of two local minimum are the same. It indicates that the HPBH/LPBH phase transition will occur at the case of the same depth for two wells from the view of Gbiss free energy landscape. At this issue, we expect that the reentrant phase transition or triple point maybe correspond to more wells of Gibbs free energy landscape.

\subsection{Fokker-Planck Equation and Probabilistic Evolution}
\label{scheme3.2}
As shown in the subsection we find that the HPBH/LPBH phase transition of the charge EPYM AdS black hole emerges when the double wells of $G_L$ have the same depth. Next we will investigate the dynamic process of the HPBH/LPBH phase transition of the charge EPYM AdS black hole.

Recently authors proposed the stochastic dynamic process of black hole phase transition can be studied by the associated probabilistic Fokker-Planch equation on Gibbs free energy landscape \cite{Li2020}, which is an equation of motion governing the distribution function of fluctuating macroscopic variables. For a black hole thermodynamic system, the horizon $r_+$ is the order parameter and it can be regarded as a stochastic fluctuating variable during phase transition. Based on this, we will exhibit the dynamical process of the HPBH/LPBH phase transition in the canonical ensemble under the thermal fluctuations. Note that the canonical ensemble is consisted of a series of black hole with arbitrary horizons. The probability distribution of these black hole states $\rho(t,r_+)$ satisfies the Fokker-Planck equation on Gibbs free energy landscape
\begin{eqnarray}
\frac{\partial\rho(t,r_+)}{\partial t}=D\frac{\partial}{\partial r_+}\left(e^{-\beta G_L(r_+)}\frac{\partial}{\partial r_+}\left[e^{\beta G_L(r_+)}\rho(t,r_+)\right]\right), \label{rho}
\end{eqnarray}
where $\beta=1/kT_E$, $D=kT_E/\xi$ is the diffusion coefficient with $k$ being the Boltzman constant and $\xi$ being dissipation coefficient. Without loss of generality, we set $k=\xi=1$. In order to solve the above equation, two types of boundary ($r_+=r_{0}$) condition should be imposed. One is the reflection boundary condition, which preserves the normalization of the probability distribution. The other is the absorption boundary condition.

In this system the location of left boundaries should be smaller than $r_s$, and the right one is bigger than $r_l$. Since the temperature of the charged EPYM AdS black hole with the pressure $P_0=0.85955P_c$ and parameters $q=0.85,~\gamma=0.8$ should be not negative, there exists the minimum value of $r_{min}=0.695808$. We can regard $r_{min}$ as the left boundary $r_{lb}$, and set the right one $r_{rb}$ as $6$. The reflection boundary condition means the probability current vanishes at the left and right boundaries:
\begin{eqnarray}
j(t,r_0)=-T_Ee^{-G_L/T_E}\frac{\partial}{\partial r_+}\left(e^{G_L/T_E}\rho(t,r_+)\right)\mid_{r_+=r_0}=0
\end{eqnarray}
And the absorption one means the vanishing probability distribution function at the boundary: $\rho(t,r_0)=0$. The adoption of boundary condition is determined by the considering physical problem.

The initial condition is chosen as a Gaussian wave packet located at $r_i$:
\begin{eqnarray}
\rho(0,r_+)=\frac{1}{\sqrt{\pi}}e^{-\frac{(r_+-r_i)^2}{a^2}}.
\end{eqnarray}
Here $a$ is a constant which determines the initial width of Gaussian wave packet and it does not influence the final result. Since we mainly consider the thermal dynamic phase transition between HPBH/LPBH states, $r_i$ can be set to $r_s$ or $r_l$. It means this thermal system is initially at high or low potential black hole state.

\begin{figure}[htp]
\centering
\subfigure[$r_i=r_l$]{\includegraphics[width=0.45\textwidth]{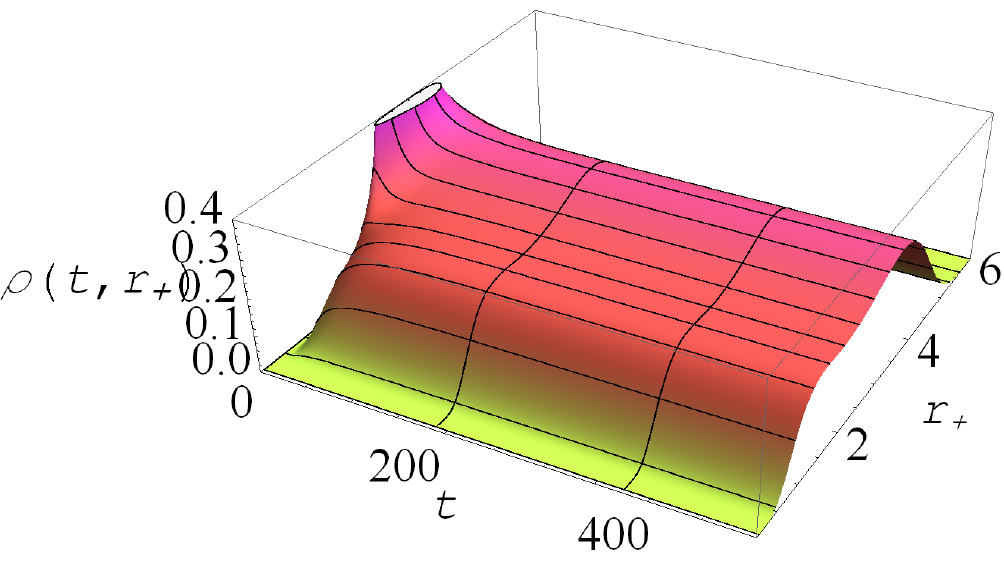}\label{rhoL8580tr}}~~~~~~
\subfigure[$r_i=r_s$]{\includegraphics[width=0.45\textwidth]{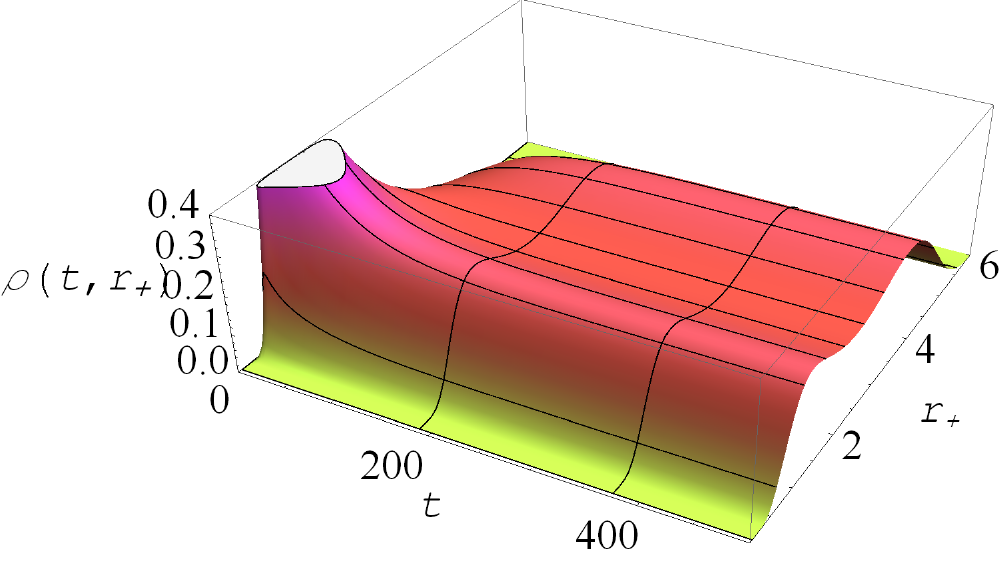}\label{rhoS8580tr}}
\caption{The plots of $\rho(t,r_+)$ at the phase transition point ($T_0=0.0402$ and $P_0=0.85955P_c$) of the EPYM AdS black holes with different initial condition and the parameters $q=0.85,~\gamma=0.8$.}\label{rho8580tr}
\end{figure}

\begin{figure}[htp]
\centering
\subfigure[$r_i=r_l$]{\includegraphics[width=0.45\textwidth]{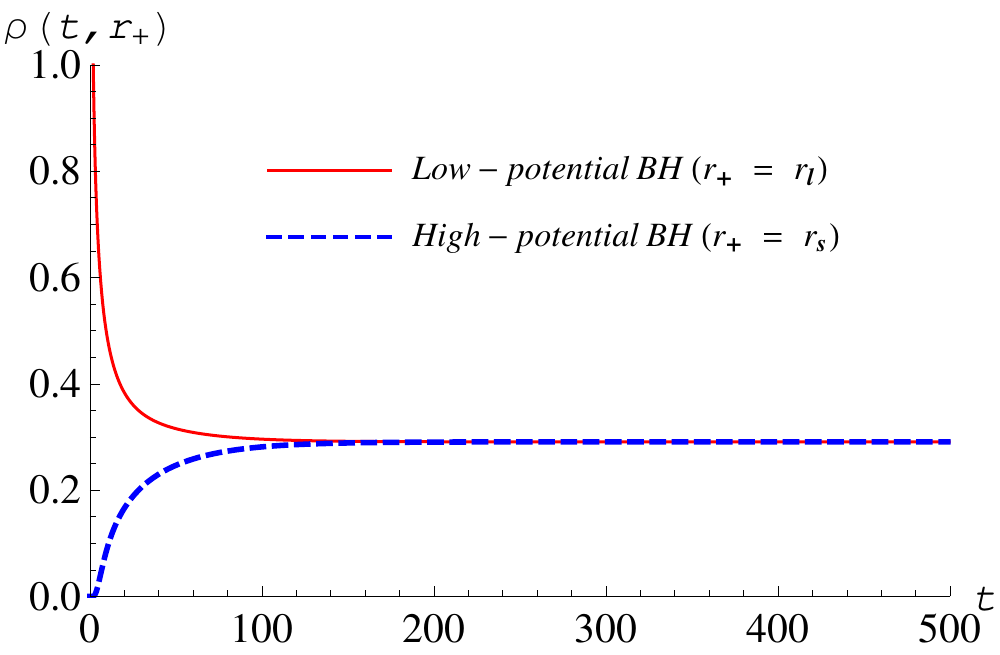}\label{rhoL8580t}}~~~~~~
\subfigure[$r_i=r_s$]{\includegraphics[width=0.45\textwidth]{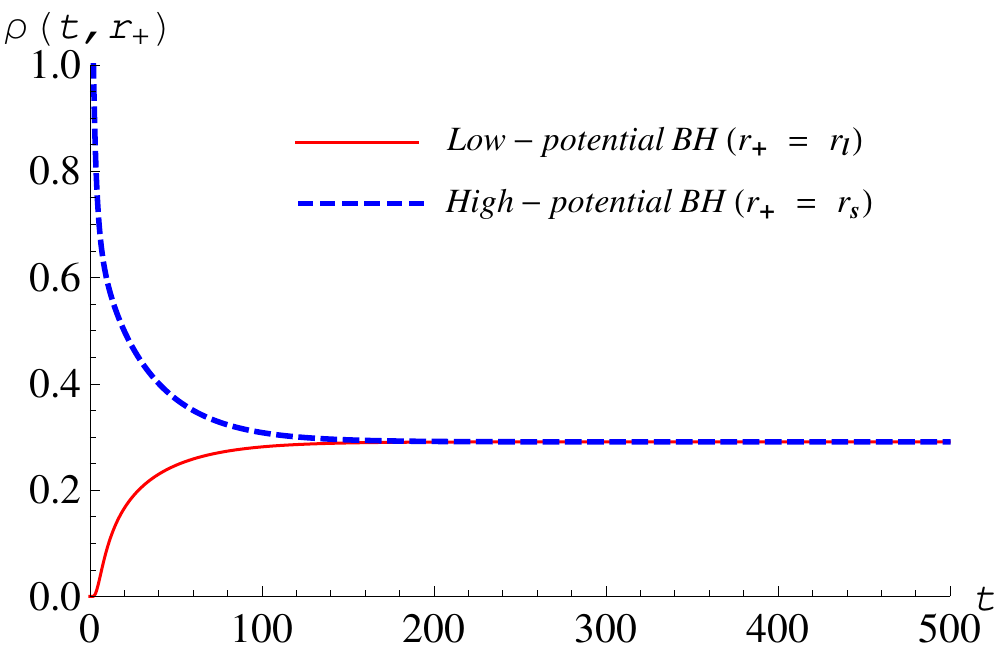}\label{rhoS8580t}}
\caption{The plots of $\rho(t)$ at the phase transition point ($T_0=0.0402$ and $P_0=0.85955P_c$) of the EPYM AdS black holes with different initial condition and the parameters $q=0.85,~\gamma=0.8$.}\label{rho8580t}
\end{figure}

The time evolution of the probability distribution are shown in Fig. \ref{rho8580tr}. As $t=0$ the Gaussian wave packets locate at LPBH (Fig. \ref{rhoL8580tr}) and HPBH (Fig. \ref{rhoS8580tr}) with $T_E=T_0$ and $a=0.1$, respectively. And they are both decreasing with increasing time $t$ until tending to a certain constant. However, at the same time the peaks of $\rho(t,r)$ at $r=r_{s}$ (Fig. \ref{rhoL8580tr}) and $r=r_{s}$ (Fig. \ref{rhoS8580tr}) are increasing from zero to the same constant. This indicates that the black hole system in LPBH phase tends to HPBH phase as shown Fig. \ref{rhoL8580tr}, while it in HPBH phase is tending to the low one as shown Fig. \ref{rhoS8580tr}. Finally the system reaches a LPBH/HPBH coexistence stationary state at a short time. In order to make the dynamical process of the HPBH/LPBH phase transition more clear, we exhibit the probability distributions of $\rho(t,r_l)$ and $\rho(t,r_s)$ in Fig. \ref{rho8580t}. The probability distribution of the initial LPBH (or HPBH) is maximum, whereas the corresponding HPBH (or LPBH) state is zero. With increasing time, both them approach to the same value. This is consistent with what has shown in $G_L-r_+$, i.e., the LPBH and HPBH states have the same depth in the double well of Gibbs free energy landscape.

\subsection{First Passage Time}
\label{scheme3.3}
In general, the important quantity in the dynamical process of the HPBH/LPBH phase transition is characterized by the first passage time, which is defined as the mean value of the first passage time that a stable HPBH or LPBH scape to the unstable intermediate-potential black hole state (i.e., from the one well to the barrier of Gibbs free energy landscape).

Supposing there is a perfect absorber in the stable HPBH or LPBH state, if the system makes the first passage under the thermal fluctuation, the system will leave this state. We can define $\Sigma$ to be the sum probability of the dynamical process within the first passage time as
\begin{eqnarray}
\Sigma=\int^{r_m}_{r_{min}}\rho(t,r_+)dr_+, ~~~~~~or~~~~~\Sigma=-\int^{r_m}_{r_{rb}}\rho(t,r_+)dr_+,\label{sigma}
\end{eqnarray}
where $r_m$, $r_{min}$, $r_{rb}$ are the intermediate, minimum, and right boundary of charge EPYM AdS black hole horizons. At a long time, the probability of the stable HPBH or LPBH state still in this system becomes zero, i.e., $\Sigma(t,r_l)\mid_{t\rightarrow\infty}=0$ or $\Sigma(t,r_s)\mid_{t\rightarrow\infty}=0$. As claimed, the first passage time is a random variable because the dynamical process of phase transition is caused by thermal fluctuation. Hence, we denote the distribution of the first passage time by $F_p$, which reads
\begin{eqnarray}
F_p=-\frac{d\Sigma}{dt}.\label{Ft}
\end{eqnarray}
It is obviously that $F_pdt$ indicates the probability of the system passing through the intermediate-potential black hole state for the first passage time in the time interval ($t,~t+dt$). Considering Eqs. (\ref{rho}) and (\ref{sigma}), the distribution of the first passage time $F_p$ becomes \cite{Li2020a}
\begin{eqnarray}
F_p=-D\frac{\partial\rho(t,r_+)}{\partial r}\mid_{r_m}, ~~~~~or~~~~F_p=D\frac{\partial\rho(t,r_+)}{\partial r}\mid_{r_m}
\end{eqnarray}
Here the absorbing and reflecting boundary conditions of the Fokker-Planck equation are imposed at $r_m$ and the other end ($r_{min}$ or $r_{rb}$). Note that the normalisation of the probability distribution is not preserved.

\begin{figure}[htp]
\centering
\subfigure[$r_i=r_l$]{\includegraphics[width=0.45\textwidth]{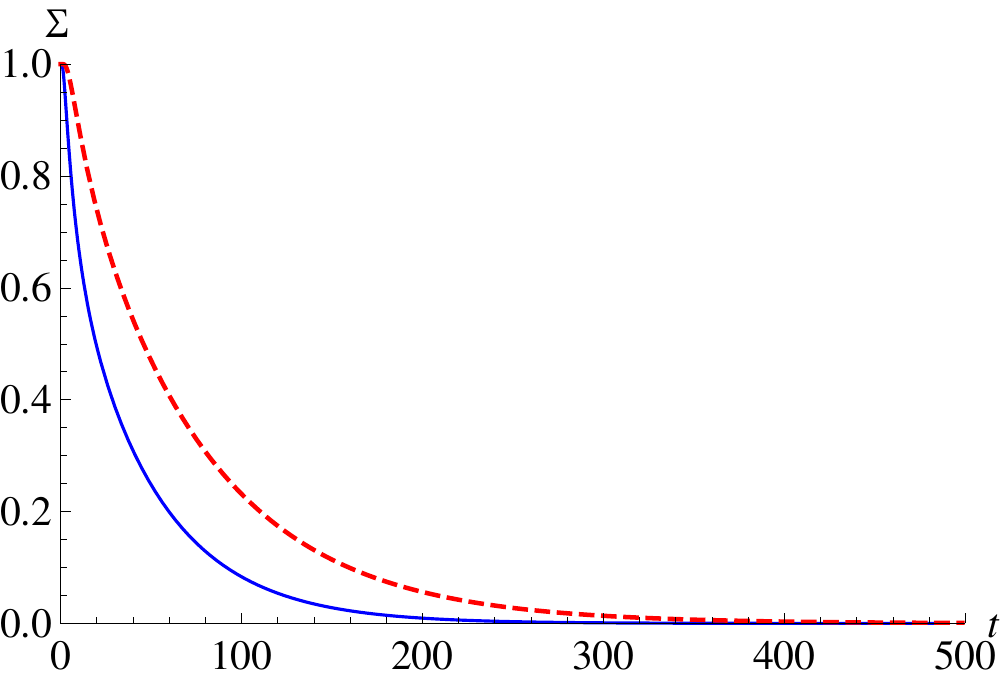}\label{sigmaL8580t}}~~~~~~
\subfigure[$r_i=r_s$]{\includegraphics[width=0.45\textwidth]{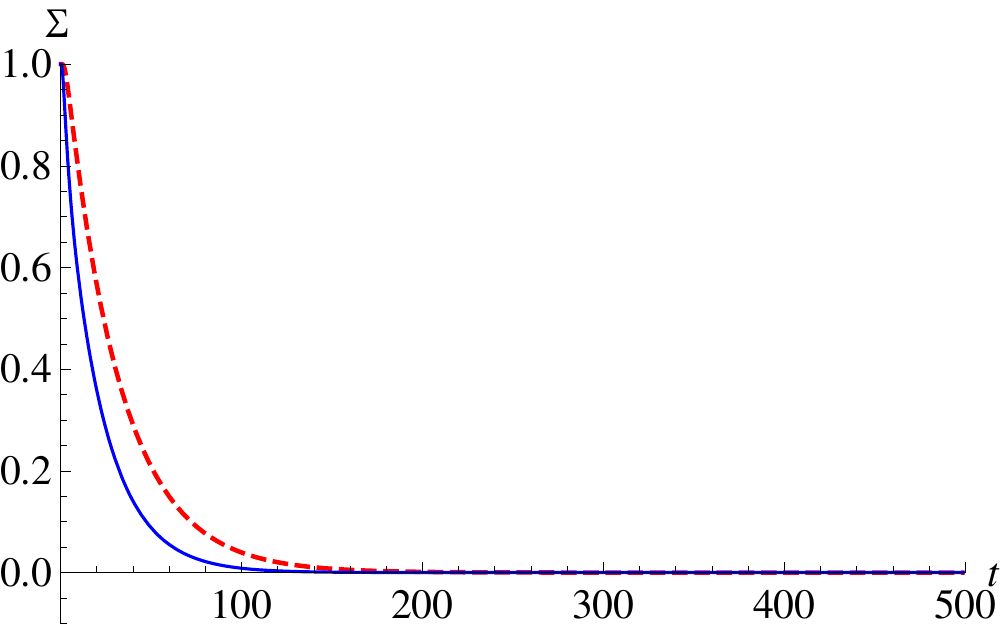}\label{sigmaS8580t}}
\caption{$\Sigma-t$ for different temperatures: $T_0=0.0402$ (thin blue lines) and $T_0=0.038$ (dashed thick red lines) for the charged EPYM AdS black holes with different initial conditions and the parameters $q=0.85,~\gamma=0.8$.}\label{sigma8580t}
\end{figure}

\begin{figure}[htp]
\centering
\subfigure[$r_i=r_l$]{\includegraphics[width=0.45\textwidth]{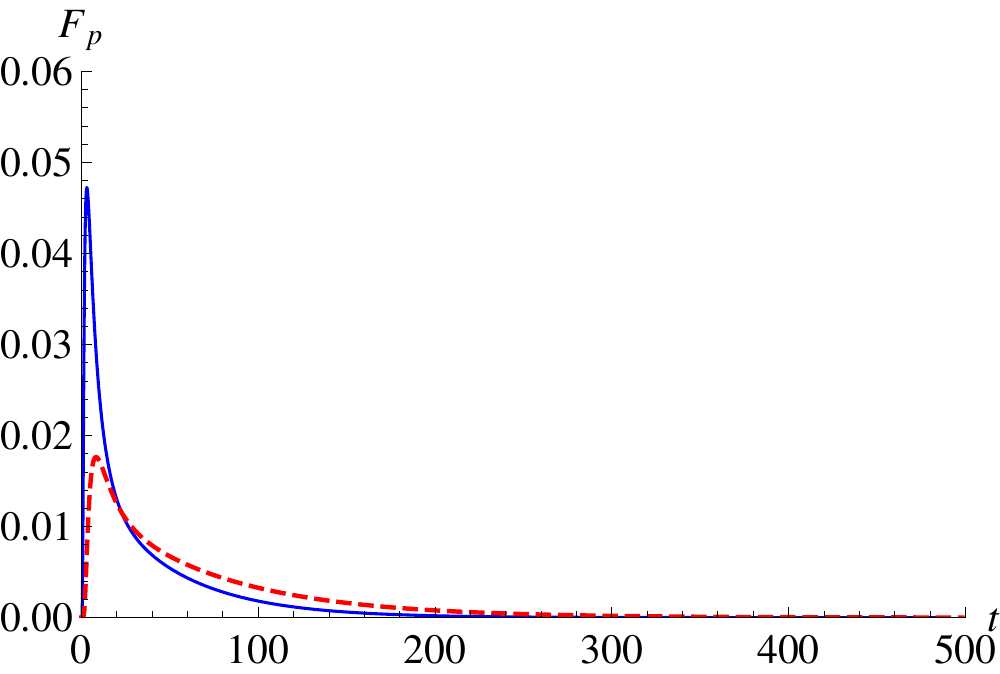}\label{Ft8580}}~~~~~~
\subfigure[$r_i=r_s$]{\includegraphics[width=0.45\textwidth]{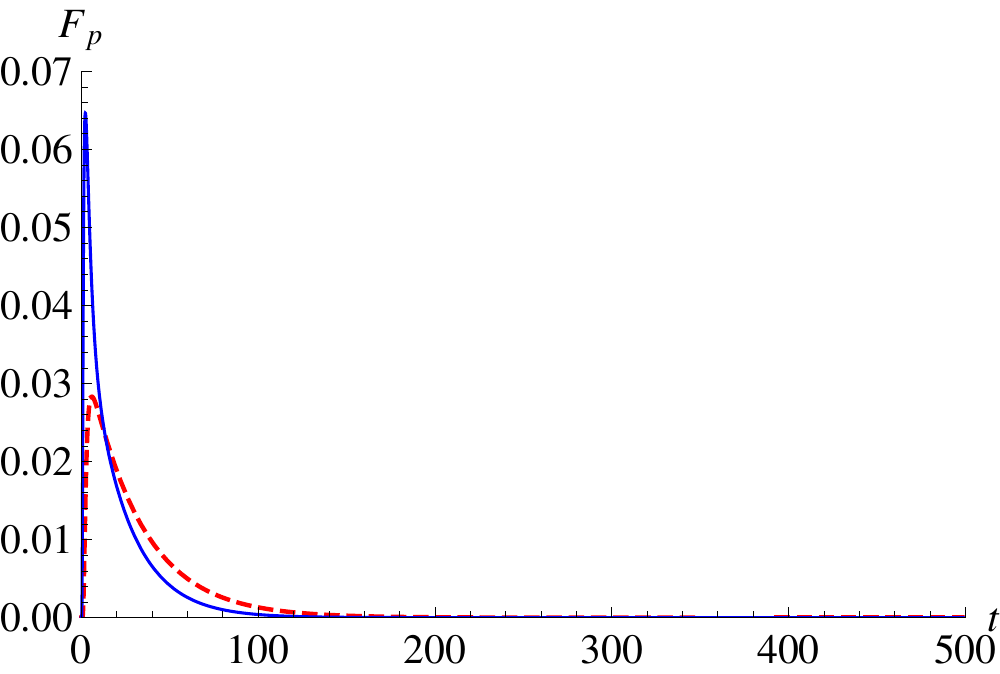}\label{Ft8580S}}
\caption{$F_p-t$ for different temperatures: $T_0=0.0402$ (thin blue lines) and $T_0=0.038$ (dashed thick red lines) for the charged EPYM AdS black holes with different initial conditions and the parameters $q=0.85,~\gamma=0.8$.}\label{Ft8580}
\end{figure}

By solving the Fokker-Planck equation (\ref{rho}) with different phase transition temperatures ($T_0=0.0402$ and $T_0=0.038$) and substituting them into Eqs. (\ref{sigma}) and (\ref{Ft}), the numerical results are displayed in Figs. \ref{sigma8580t} and \ref{Ft8580} for different initial conditions. It is obviously clear that in Fig. \ref{sigma8580t}, no mater what kind initial condition has been considered $\Sigma$ decays very fast in a short time. Furthermore the increase of temperature makes the probability of $\Sigma$ drop faster. An important point to pay is that the probability is not conserved. From the corresponding probability distribution picture in Fig. \ref{Ft8580}, the behavior of $F_p$ is the similar for both two different initial conditions. A single peak emerges near $t=0$ in the curve of $F_p$ with fixed temperature. This can be understood as a large number of first passage events occur in a short interval of time, and the probability distribution decays exponentially with time. The effect of temperature at phase transition points on $F_p$ is consistent with that of $\Sigma$ and $G_L$. That means the higher temperature, the faster probability decreases, the easier phase transition occurs, and the lower depth of the barrier is; otherwise the lower temperature, the slower probability decreases, the harder phase transition occurs, and the higher depth of the barrier is.
\section{Discussions and Conclusions}
\label{scheme4}
Since the precise statistical description of the corresponding thermodynamic states of black holes is still unclear, the investigation of the thermodynamic phase transition of black holes becomes a concerned issue. In this manuscript, we investigated the dynamical property of the HPBH/LPBH phase transition for the four-dimensional charge Einstein-power-Yang-Mills (EPYM) AdS black hole from the view point of Gibbs free energy landscape.

Firstly we reviewed the thermodynamic property of the charged EPYM AdS black hole. From the phase transition condition (\ref{PTC}), we suggested that the phase transition is the HPBH/LPBH one, while is not only the pure one between a small/big black hole. As what we had presented, the HPBH/LPBH phase transition in $T-S$ diagram can be constructed when $P_0<P_c$ by Maxwell's equal-area law. Since the results are the same no mater from Maxwell's equal-area law or Gibbs free energy, the swallow tail behavior of $G$ was not presented in this manuscript.

At the phase point what we had given, we found there is the double-well in Gibbs free energy landscape ($G_L$). The two local minimum of $G_L$ correspond to the stable HPBH/LPBH states. The local maximum stands for the unstable intermediate-potential black hole state and acts as a barrier between the stable HPBH and LPBH states. Furthermore, the depth of two wells are the same. It indicates that the HPBH/LPBH phase transition will occur when two wells have the same depth from the point view of $G_L$. Next we studied the dynamical process of the HPBH/LPBH phase transition governed by the Forkker-Planck equation. By imposing the reflection boundary conditions on the minimum black hole horizon and the merely bigger value than the LPBH, and considering a Gaussian wave packet at the HPBH or LPBH state as the initial condition, we obtained the numerical result of the Forkker-Planck equation: the initial Gaussian wave packet at the HPBH or LPBH state decreases with increasing with time, however at the same time the other peak of $\rho(t,r_+)$ at the LPBH or HPBH state increases from zero to a same constant. That indicates that with increasing time the system will leave from the initial state to another state, until it becomes a two-state coexistent state, which is consistent with the fact that the depth of two wells of $G_L$ (standing for the LPBH and HPBH states) are the same value.

Finally we considered the first passage time. By imposing the absorption boundary condition on the intermediate-potential black hole state and considering a Gaussian wave packet at the HPBH or LPBH state as the initial condition, we also obtained the numerical result of the Forkker-Planck equation: no mater what kind initial condition had been considered $\Sigma$ decays very fast in a short time and it drops faster with increasing temperature. And the behavior of $F_p$ is the similar for both two different initial conditions. There exists a single peak near $t=0$ in $F_p$. This can be understood as a large number of first passage events occur in a short interval of time, and the probability distribution decays exponentially with time. From the effect of temperature at phase transition points on $F_p$, $\Sigma$, and $G_L$, we found the higher temperature, the faster probability decreases, the easier phase transition occurs, and the lower depth of the barrier is; otherwise the lower temperature, the slower probability decreases, the harder phase transition occurs, and the higher depth of the barrier is.

\section*{Acknowledgements}
We would like to thank Prof. Zong-Hong Zhu and Meng-Sen Ma for their indispensable discussions and comments. This work was supported by the National Natural Science Foundation of China (Grant No. 11705106, 11475107, 12075143), the Natural Science Foundation of Shanxi Province, China (Grant No.201901D111315), the Natural Science Foundation for Young Scientists of Shanxi Province, China (Grant No.201901D211441), the Scientific Innovation Foundation of the Higher Education Institutions of Shanxi Province (Grant Nos. 2020L0471, 2020L0472, 2016173), and the Science Technology Plan Project of Datong City, China (Grant Nos. 2020153).

\end{document}